\documentclass[
superscriptaddress
,preprint
,tightenlines
,showpacs
,showkeys
,nofootinbib
,aps
,amsfonts
,amssymb
,byrevtex
,pra,
,pdflatex 
]{revtex4-1}

\usepackage{graphicx} 
\usepackage{amsmath, amssymb, bm} 
\usepackage{braket}
\usepackage{subcaption}
\usepackage{ragged2e}
\DeclareCaptionJustification{justified}{\justifying}
\usepackage{multirow}

\def\){\right)} 
\def\({\left(} 
\def\]{\right]} 
\def\[{\left[}

\def\affHISKP{\affiliation{Helmholtz-Institut f\"ur Strahlen- und Kernphysik~(Theorie) \\ 
and Bethe Center for Theoretical Physics, Universit\"at Bonn, D-53115 Bonn, Germany}}
\def\affKMU{\affiliation{Department of Physics, Karamanoglu Mehmetbey University, Karaman 70100, Turkey}}
\def\affNCSU{\affiliation{Department of Physics, North Carolina State University, \\
Raleigh, North Carolina 27695, USA}}
\def\affJUELICH{\affiliation{Institute for Advanced Simulation, Institut f{\"u}r Kernphysik, \\
J{\"u}lich Center for Hadron Physics and JARA - High Performance Computing, \\ Forschungszentrum J{\"u}lich, D-52425 J{\"u}lich, Germany}}

\def\affMSU{\affiliation{Department of Physics $\&$ Astronomy and HPC$^2$ Center for Computational Sciences, 
Mississippi State University, Mississippi State, Mississippi State 39762, USA}}

\begin{document}


\title{Universal dimer-dimer scattering in lattice effective field theory}

\author{Serdar~Elhatisari}
\email{elhatisari@hiskp.uni-bonn.de}
\affHISKP
\affKMU

\author{Kris~Katterjohn}
\email{kk278@msstate.edu}
\affMSU

\author{Dean~Lee}
\email{dean\_lee@ncsu.edu}
\affNCSU

\author{Ulf-G.~Mei{\ss}ner}
\email{meissner@hiskp.uni-bonn.de}
\affHISKP
\affJUELICH

\author{Gautam~Rupak}
\email{grupak@u.washington.edu}
\affMSU

\begin{abstract}
We consider two-component fermions with short-range interactions and large scattering length. This system has universal properties that are realized in several different fields of physics. In the limit of large fermion-fermion scattering length $a_\mathrm{ff}$ and zero-range interaction, all properties of the system scale proportionally with $a_\mathrm{ff}$. For the case with shallow bound dimers, we calculate the dimer-dimer scattering phase shifts using lattice effective field theory. We extract the universal  dimer-dimer scattering length $a_\mathrm{dd}/a_\mathrm{ff}=0.618(30)$ and effective range $r_\mathrm{dd}/a_\mathrm{ff}=-0.431(48)$.  This result for the effective range is the first calculation with quantified and controlled systematic errors.  We also benchmark our methods by computing the fermion-dimer scattering parameters and testing some predictions of conformal scaling of irrelevant operators near the unitarity limit.

\end{abstract}

\date{\today}
\maketitle


\section{Introduction}
\label{sec_intro}

Two-component fermions at large scattering length are an important system with universal properties and relevance to several branches of physics.  This universality is due to the existence of a conformal fixed point called the unitarity limit where the fermion-fermion scattering length is infinite and all other length scale are irrelevant at large particle separations or low energies.  See for example Ref.~\cite{Braaten:2004rn} for a review.
In nuclear physics, the neutron-neutron scattering length $|a_{nn}|\sim 19$ fm~\cite{GonzalezTrotter:1999zz} is much larger than the inverse pion mass $1/M_\pi\sim 1.4$ fm characterizing the exponential tail of the nuclear force.  In the physics of ultracold atoms, one can tune the interactions arbitrarily close to the unitarity limit
using an external magnetic field near a Feshbach resonance~\cite{Feshbach:1962ut,Inouye:1998}.  In this letter we discuss the case where the scattering length is large and positive, and bound dimers composed of two fermions are formed with shallow binding energy.  We compute dimer-dimer scattering and determine the the dimer-dimer scattering length and effective range.  These results can be used to compute the energy density of a dimer gas in the dilute limit \cite{Zwerger:2011,Lee:1957a,Lee:1957b,Wu:1959,Braaten:1999}.

The elastic scattering phase shift $\delta(p)$ between two non-relativistic fermions with finite-range interactions is parameterized by the effective range expansion (ERE) \cite{Bethe:1949yr},
\begin{align}
p\cot\delta=-\frac{1}{a_\mathrm{ff}}+\frac{1}{2} \, r_\mathrm{ff} \, p^2+\mathcal O(p^4)\,,
\label{eq:ERE}
\end{align}
where $p$ is the relative momentum, $a_\mathrm{ff}$ is the fermion-fermion scattering length, and $r_\mathrm{ff}$ is the fermion-fermion effective range. In this study we consider the case where  $a_\mathrm{ff}$ is large and positive while all other lengths scales are negligible.  We can express all physical quantities in dimensionless combinations involving powers of $a_\mathrm{ff}$.  Previous calculations of the dimer-dimer scattering length have found $a_\mathrm{dd}/a_\mathrm{ff}=0.60\pm0.01$~\cite{Petrov:2004zz,Petrov:2005zz}, $a_\mathrm{dd}/a_\mathrm{ff}=0.605\pm0.005$~\cite{Greene:2009}, and $a_\mathrm{dd}/a_\mathrm{ff}=0.60$~\cite{Aurel:2004}. A perturbative expansion about four spatial dimensions gives $a_\mathrm{dd}/ a_\mathrm{ff}\approx0.66$~\cite{Rupak:2006jj}, and a rough estimate using the resonating group method in the single-channel approximation
gives $a_\mathrm{dd}/a_\mathrm{ff}\sim0.752$ \cite{Naidon:2016}. 
On the other hand, much less
is known about the higher-order dimer-dimer ERE parameters. The effective range has been calculated as $r_\mathrm{dd}/a_\mathrm{ff} \approx 0.12$ in Ref.~\cite{vonStecher:2007zz}, while a very rough estimate of $r_\mathrm{dd}/a_\mathrm{ff} \sim 2.6$ was given in Ref.~\cite{Lee:2007ae}.

In this work we calculate the low-energy dimer-dimer phase shifts from lattice effective field theory  and extract both the dimer-dimer scattering length $a_\mathrm{dd}$ and effective range $r_\mathrm{dd}$. We also benchmark our methods by calcuating the fermion-dimer scattering length
$a_\mathrm{fd}$ and effective range $r_\mathrm{fd}$.  We organize our paper as follows. In Sec.~\ref{sec_formalism} we introduce the continuum and lattice formulations for systems of two-component fermions. In Sec.~\ref{sec_phaseshift} we discuss the methods for extracting the scattering information from periodic finite volumes. We present our results and analyses for fermion-dimer and dimer-dimer scattering in Sec.~\ref{sec_results}. The results are summarized in Sec.~\ref{sec_conclusion}.

\section{Lattice Formalism}
\label{sec_formalism}

Following Refs.~\cite{Lee:2008fa,Elhatisari:2016hby}, we start by describing interacting two-component fermions in continuous space. Low-energy fermion-fermion scattering
is dominated by the $s$-wave channel, while higher partial waves become more important at higher energies. In principle the two components could have different masses, however we only consider the equal mass case in this study and denote the two components as up and
down spins. We will consider systems of two-component fermions with different masses in a future publication. 

We work with natural units where $\hbar=1=c.$ Let $b^{\,}_{\uparrow,\downarrow}$ ($b_{\uparrow,\downarrow
  }^{\dagger}$) be the annihilation (creation) operators, and let $\rho_{\uparrow,\downarrow}$ be the density operators,
\begin{equation}
\rho^{\,}_{\uparrow}(\vec{r})
= b^{\dagger}_{\uparrow}(\vec{r}) b^{\,}_{\uparrow}(\vec{r}), \qquad \rho^{\,}_{\downarrow}(\vec{r})
= b^{\dagger}_{\downarrow}(\vec{r}) b^{\,}_{\downarrow}(\vec{r})\,.
\label{eqn:density-operator-001}
\end{equation}
The continuum Hamiltonian has the form
\begin{equation}
H = \sum_{s=\uparrow,\downarrow}
\frac{1}{2m}
\int d^{3}\vec{r} \, 
\vec{\nabla}b^{\dagger}_{s}(\vec{r})
\cdot
\vec{\nabla}b^{\,}_{s}(\vec{r})
+ C_{0} \int d^{3}\vec{r}  \, 
\rho_{\uparrow}(\vec{r}) \rho_{\downarrow}(\vec{r})
\,,
\label{eqn:Hamiltonian-001}
\end{equation} 
where ultraviolet divergences due to the zero-range interaction are regulated
in some manner. In our case we use the lattice to provide the ultraviolet regularization.

We denote the lattice spacing as $a$. In our calculations we use an $\mathcal{O}(a^{4})$-improved lattice action where the free lattice Hamiltonian, $H_\mathrm{0}$, is defined as,
\begin{align}
H_{0} 
= \sum_{s=\uparrow,\downarrow}
\,
\frac{1}{2m}
\,
\sum_{\hat{l}=\hat{1},\hat{2},\hat{3}}
\,
\sum_{\vec{n}}
\, 
\[
\sum_{k=-3}^{3}\, w_{|k|} \, 
b^{\dagger}_{s}(\vec{n})b^{\,}_{s}(\vec{n}+k\,\hat{l})
\],
\label{eqn:free-Hamiltonian-001}
\end{align} 
where $\hat{l}=\hat{1},\hat{2},\hat{3}$ are unit vectors in spatial directions, and the hopping parmeters $w_{0}$, $w_{1}$, $w_{2}$, and $w_{3}$ are 49/18, $-$3/2, 3/20, and $-$1/90, respectively. $\vec{n}$ denotes the lattice sites on a three-dimensional $L\times L\times L$ periodic
cube.

For the two-particle (2$N$) interaction we use the single-site interaction
\begin{align}
V_{2N} 
= C_{2N} \, 
\sum_{\vec{n}}
\rho_{\uparrow}(\vec{n}) \, \rho_{\downarrow}(\vec{n})
\,,
\label{eqn:Interaction-001}
\end{align} 
where the value of $C_{2N}$ depends on the lattice spacing $a$. We tune $C_{2N}$  to produce the desired value of the dimer binding energy
$B_{\mathrm{d}}$.
For convenience we choose parameters
typical for nuclear physics, with fermion mass $m = 939$ MeV and dimer binding energies ranging from $1$ MeV to $10$ MeV. However the final
results are completely independent of these details when expressed in terms of the
two-fermion scattering length $a_\mathrm{ff}$.

In the low-energy limit of this theory, three-particle and higher-particle interactions are irrelevant operators. Nevertheless, we find it useful to include three-particle $(3N)$ and four-particle $(4N)$ interactions as a diagnostic tool to generate more data for the continuum-limit extrapolations.  The three-particle interaction we use features nearest-neighbour and next-to-nearest-neighbour interactions,
\begin{align}
V_{3N}
= C^{(1)}_{3N} &  
\sum_{\vec{n}}
\sum_{|\vec{n}-\vec{n}'|^2=1}
\rho_{\uparrow}(\vec{n}) \, \rho_{\downarrow}(\vec{n}) \,
[\rho_\uparrow(\vec{n}')+\rho_\downarrow(\vec{n}')]
\nonumber\\
& + C^{(2)}_{3N}  
\sum_{\vec{n}}
\sum_{|\vec{n}-\vec{n}'|^2=2}
\rho_{\uparrow}(\vec{n}) \, \rho_{\downarrow}(\vec{n}) \,
[\rho_\uparrow(\vec{n}')+\rho_\downarrow(\vec{n}')] \, .
\label{eqn:3NInteraction}
\end{align}
Similarly, we introduce a four-particle interaction that consists of nearest-neighbour and next-to-nearest-neighbour interactions,
\begin{align}
V _{4N}
= C^{(1)}_{4N} &
\sum_{\vec{n}}
\sum_{|\vec{n}-\vec{n}'|^2=1}\rho_{\uparrow}(\vec{n}) \, \rho_{\downarrow}(\vec{n}) \,
\rho_{\uparrow}(\vec{n}') \, \rho_{\downarrow}(\vec{n}')
\nonumber\\
& + C^{(2)}_{4N}\sum_{\vec{n}}
\sum_{|\vec{n}-\vec{n}'|^2=2}\rho_{\uparrow}(\vec{n}) \, \rho_{\downarrow}(\vec{n})
\,
\rho_{\uparrow}(\vec{n}') \, \rho_{\downarrow}(\vec{n}')
\,.
\label{eqn:4NInteraction}
\end{align} 
We keep the interaction strengths $C^{(1,2)}_{3N}$ and $C^{(1,2)}_{4N}$ at fixed values when measured in lattice units.  

We can compute the importance of these irrelevant operators in the continuum limit near the conformally-invariant point where the two-fermion scattering length is infinite and the interaction range is zero.  If $\delta$ is the scaling dimension of an operator $O$,  then the contribution from the insertion of the interaction $O^{\dagger}O$ scales as $a^{2\delta-d-2}$ in the continuum limit, where $d$ is the number of spatial dimensions \cite{Nishida:2010tm}. The operator-state correspondence principle connects the scaling dimension of an operator to the lowest energy of the system in a harmonic trap with the corresponding number of  particles and quantum numbers \cite{Nishida:2007pj}.  From numerical calculations of the harmonically-trapped energies, we deduce that the leading behavior of the three-particle operators is $a^{3.54544}$, while the leading behavior
of the four-particle operators is $a^{5.056}$ \cite{Nishida:2010tm}. In our analysis we will check explicitly if this dependence on the lattice spacing can be seen in the lattice results.

\section{Scattering phase shift}
\label{sec_phaseshift}

L\"uscher's finite-volume method relates the two-body energy levels in a cubic periodic box to the elastic scattering phase shifts~\cite{Luscher:1986pf,Luscher:1991ux}. The two-body phase shifts in a periodic box of size $L$ are related to the relative momentum of the two bodies, $p$, by the relation 
\begin{align}
p\cot\delta(p)=\frac{1}{\pi L}\,S(\eta),\ \ \eta=\(\frac{p \, L}{2\pi}\)^2,
\label{eqn:Luescher-001}
\end{align} 
where $S(\eta)$ is the regulated three-dimensional zeta function given by
\begin{align}
S(\eta)=\lim_{\Lambda\rightarrow\infty}\[\sum_{\vec{n}}
\frac{\Theta(\Lambda^2-\vec{n}^2)}{\vec{n}^2-\eta}-4\pi\Lambda
\].
\label{eqn:Zeta_function-001}
\end{align}
The sum in Eq.~(\ref{eqn:Zeta_function-001}) is over three-dimensional integer vectors $\vec{n}$. 

We use the Lanczos eigenvector method \cite{Lanczos:1950} to compute the low-energy spectrum of the lattice Hamiltonian at different values of $L$.  These energies levels determine the values of $p$ as input into Eq.~(\ref{eqn:Luescher-001}), which then determine the two-body scattering phase shifts $\delta(p)$.
First we do these calculations for the three-particle system to determine the fermion-dimer scattering parameters.  We then do the calculations for the four-particle system to determine the
dimer-dimer scattering parameters.

\begin{figure}[!ht]
\includegraphics[width=0.5\textwidth]{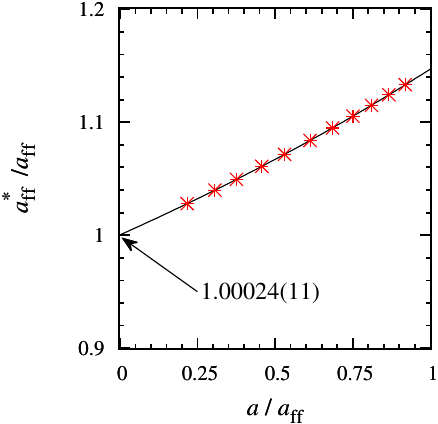}
\caption{The ratio of the fermion-fermion scattering length $a^*_\mathrm{ff}$ determined using L{\"u}scher's
finite volume formula and $a_\mathrm{ff}$ determined from $B_{\rm d}$.  The results are plotted versus the lattice spacing $a$ as a fraction of $a_\mathrm{ff}$, and fitted to a polynomial in $a/a_{\rm ff}$.}
\label{fig:affLattice-continuum}
\end{figure}

In the zero-range limit, the fermion-fermion scattering length is related to the dimer binding energy by the formula 
\begin{equation}
B_{\mathrm{d}}=\frac{1}{m a_\mathrm{ff}^2}. \label{eqn:Bd_aff}
\end{equation} 
Since we will take the zero-range limit in all our calculations, we can define $a_\mathrm{ff}$ quite simply using the zero-range formula in Eq.~(\ref{eqn:Bd_aff}) and the dimer binding energy $B_{\mathrm{d}}$ determined on the lattice.  However, we can also determine the fermion-fermion scattering length more carefully using L\"uscher's finite-volume scattering method. We call this determination of the scattering length $a^*_\mathrm{ff}$. In Fig.~\ref{fig:affLattice-continuum} we show the ratio of the scattering lengths for various lattice spacings $a$.  In the plot we have fitted a polynomial in $a/a_{\rm ff}$ to the results. We see that the 
deviation between these two definitions of the scattering length vanishes in the continuum limit and  can be fit well by a polynomial in $a/a_{\rm ff}$.

We will use L\"uscher's finite-volume
method to calculate fermion-dimer scattering and dimer-dimer scattering. In these cases we consider the scattering of two bodies, where one or both  bodies may be dimers. Let $\mu_{12}$ be the reduced mass of the two scattering objects.   In the infinite volume and continuum limits, the relative momentum $p$ is related to the two-body energy level $E^{(\infty)}$ as
\begin{align}\label{eq:Edd_infinity}
E^{(\infty)}=\frac{p^2}{2\mu_{12}}
-B_{1}-B_{2},
\end{align} 
where $B_1$ and $B_2$ are the infinite-volume binding energies for the two bodies.  These will equal $B_{\rm d}$ if a dimer or $0$ if a fermion.   
 
 Eq.~(\ref{eq:Edd_infinity}) is modified by several effects at finite volume and nonzero lattice spacing. At nonzero lattice spacing, the effective mass of the  dimer is not exactly equal to twice the fermion mass.  So we  numerically calculate the energy versus momentum dispersion relation of the dimer to extract the dimer effective mass.  For this we use a large $L = 50a$ cubic box in order to minimize finite-volume errors. From the dimer effective mass we can determine the reduced mass of the fermion-dimer and dimer-dimer systems.  We write this lattice-determined reduced mass as $\mu^*_{12}$. 

At finite volume, there is also a finite-volume correction to the binding energies $B_1$ and $B_2$. These finite-volume corrections vanish exponentially with the size of the box and so can be ignored for sufficiently large $L$.  However computational limits often make very large volume calculations impractical, and it so is  useful to remove finite-volume corrections corresponding to binding energies when possible.  It turns out that the finite-volume corrections to the binding energies $B_1$ and $B_2$ are momentum dependent. We account for these finite-volume momentum-dependent effects  using finite-volume topological factors $\tau(\eta)$ due to the dimer wave functions wrapping around the periodic box~\cite{Bour:2011ef}, where $\eta$ was defined in Eq.~(\ref{eqn:Luescher-001}). With these corrections, Eq.~(\ref{eq:Edd_infinity}) becomes
\begin{align}\label{eq:E12_L}
E^{(L)}=\frac{p^2}{2\mu^\ast_{12}}
-B^{\,}_{1} - \tau_{1}(\eta)\,\Delta B^{(L)}_{1}
-B^{\,}_{2} - \tau_{2}(\eta)\,\Delta B^{(L)}_{2} \,,  
\end{align}
where $\Delta B^{(L)}_{i}$ is the finite-volume correction $\Delta B^{(L)}_{i} = B^{(L)}_{i}-B_{i}$. The topological factor is given by~\cite{Bour:2011ef}
\begin{align}\label{eq:taud}
\tau(\eta) = &
\[{\sum_{\vec{k}}\frac{1}{(\vec{k}^{2}-\eta)^2}}\]^{-1}\sum_{\vec{k}}
\frac{\sum_{i=1}^3\cos(2\pi\alpha\,k_i)}{3(\vec{k}^2-\eta)^{2}} \,,
\end{align}
where $\vec{k}$ runs over all integer vectors, and $\alpha = 1/2$ for the dimer bound state. The relative momentum $p$ corresponding to box length $L$ is computed by solving Eq.~(\ref{eq:E12_L}) self-consistently for the given lattice energies $E^{(L)}$, $B^{(L)}_{1,2}$, and $B_{1,2}$.

\section{Results and Analysis}
\label{sec_results}

\subsection{Fermion-dimer scattering}
\label{sec-Fermion-dimer scattering}
Before proceeding to the dimer-dimer system, we perform benchmarks of our lattice methods and analysis by computing fermion-dimer scattering. Fermion-dimer scattering
has been calculated using semi-analytical methods~\cite{Bedaque:1997qi,Bedaque:1998mb,Gabbiani:1999yv,Bedaque:2002yg}
in the continuum limit. We consider a three-particle system of two spin-up fermions and one spin-down fermion. Our lattice Hamiltonian
has the form
\begin{equation}
H = H_0 + V_{2N} + V_{3N},
\end{equation}
where the free Hamiltonian is defined in Eq.~(\ref{eqn:free-Hamiltonian-001}), the two-particle interaction
appears in Eq.~(\ref{eqn:Interaction-001}),
and the three-particle interaction is introduced
in Eq.~(\ref{eqn:3NInteraction}).  In order to reduce the number of free parameters in our analysis, we define the three-particle parameter $c_{3N}$ and set 
$C^{(1)}_{3N}=c_{3N}$
and $C^{(2)}_{3N}=c_{3N}/2$.
In the following we quote the value of $c_{3N}$ in lattice units.

\begin{figure}[!ht]
\includegraphics[width=0.80\textwidth]{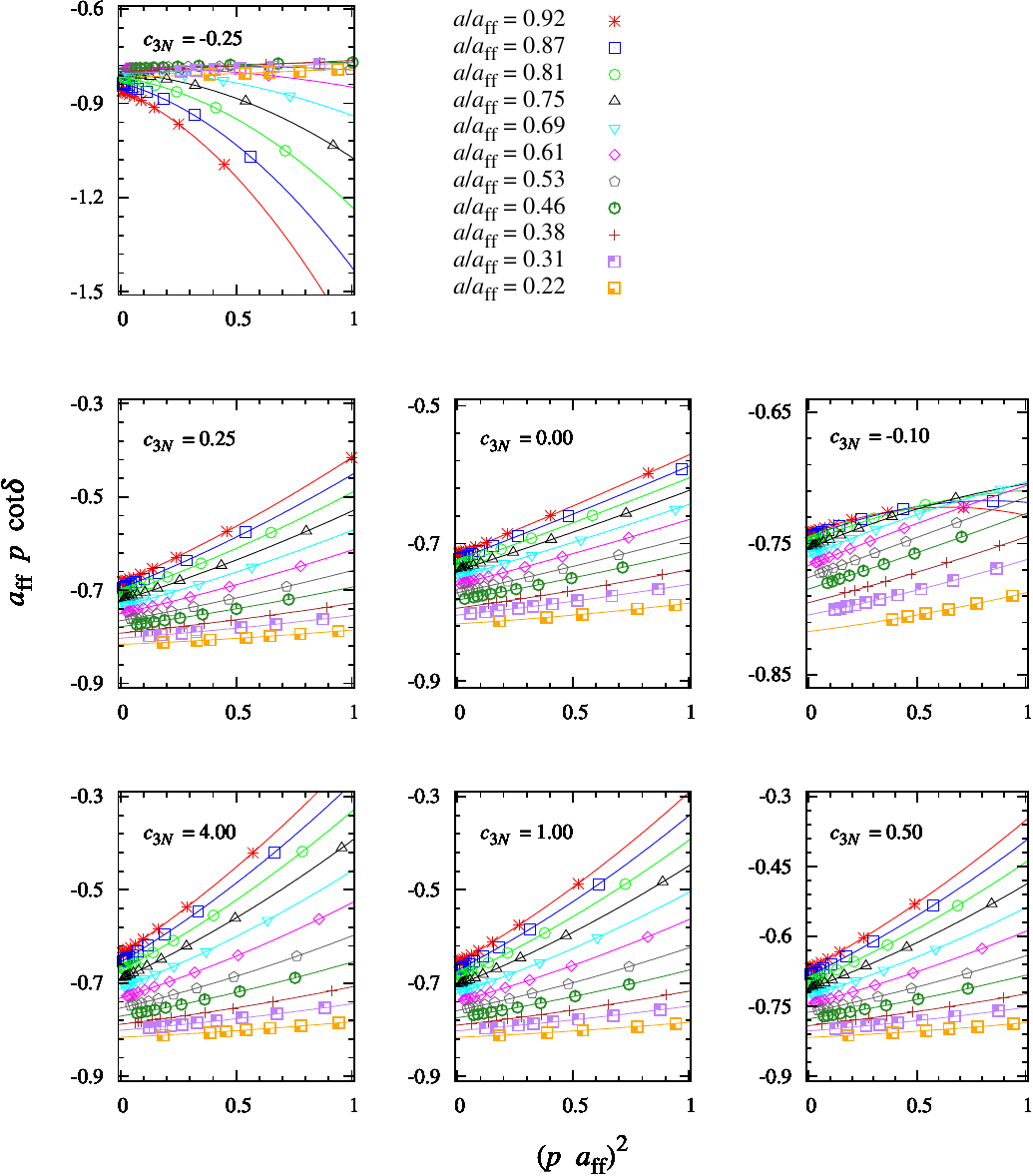}
\caption{The fermion-dimer scattering results for various values of the three-particle coupling $c_{3N}$ in lattice units and various ratios of the lattice spacing $a$ to the fermion-fermion scattering length $a_{\rm ff}$. We plot $a_{\mathrm{ff}}\,p\cot\delta$ versus $(a_{\mathrm{ff}}\,p)^{2}$ in the center-of-mass frame. The points are the lattice data, and the lines are the fits to the effective range expansion.}
\label{fig:fermidimer_phaseshifts}
\end{figure}

We perform lattice calculations using the
Lanczos eigenvector method to obtain the finite-volume energies of the fermion-dimer system for various interaction coefficients $C_{2N}$ and $c_{3N}$ and lattice lengths $L$. From the fermion-dimer energies 
$E^{(L)}_{\mathrm{fd}}$ in the center-of-mass frame, we determine the relative momentum $p$ using
\begin{align}\label{eq:Efd_L}
E^{(L)}_{\mathrm{fd}} = \frac{p^2}{2\mu^{\ast}_{\mathrm{fd}}}
-B^{\,}_{\mathrm{d}} - \tau_{\mathrm{d}}(\eta) \, \Delta B^{(L)}_{\mathrm{d}} \,, 
\end{align}
and then use Eq.~(\ref{eqn:Luescher-001}) to extract the fermion-dimer scattering phase shifts. 

The results for the fermion-dimer phase shifts are shown in Fig.~\ref{fig:fermidimer_phaseshifts}. We plot $a_{\mathrm{ff}}\,p\cot\delta$ versus $(a_{\mathrm{ff}}\,p)^{2}$
for various values of the three-particle
coupling $c_{3N}$ and various ratios of the lattice spacing
$a$ to the fermion-fermion scattering length $a_{\rm ff}$.  The values quoted for $c_{3N}$ are in lattice units.  In each case we make a fit using the truncated effective range expansion 
\begin{equation}
a_{\mathrm{ff}}\,p\cot\delta =- \frac{1}{a_{\rm fd}/a_{\rm ff}} + \frac{1}{2}r_{\rm fd}/a_{\rm ff} \cdot (a_{\rm ff}\,p)^2 + O(p^{4}) ,
\end{equation}
where $a_{\rm fd}$ and $r_{\rm fd}$ are the fermion-dimer scattering length and  effective range respectively.  As seen in Fig.~\ref{fig:fermidimer_phaseshifts}, the three-particle interactions have some impact on the scattering phase shift results at larger lattice spacings, while the data at small $a$ is almost independent of $c_{3N}$.  This is consistent with the conformal scaling prediction that the three-particle interactions are irrelevant in the continuum limit.

With these lattice results for $a_{\rm fd}$ and $r_{\rm fd}$, we extrapolate to the continuum limit.  There will be lattice cutoff corrections that scale as integer powers of the lattice spacing. For these corrections we fit a third-order polynomial in $a/a_{\rm ff}$ with coefficients that are independent of  $c_{3N}$.  We also include the predicted leading order correction from $c_{3N}$ as $(a/a_{\rm ff})^{3.54544}$  as well as a subleading correction at one power higher, $(a/a_{\rm ff})^{4.54544}$. We could include other corrections as well, however there is a limit to the number of powers that can be fit reliably at the same time. In summary, we perform the continuum-limit extrapolations for $a_{\rm fd}$ and $r_{\rm fd}$ using the functional form
\begin{align}
f({a}/{a_{\mathrm{ff}}}) & = f_{0}
       + f_{1} \, ({a}/{a_{\mathrm{ff}}})  
       + f_{2} \, ({a}/{a_{\mathrm{ff}}})^{2}
        + f_{3} \, ({a}/{a_{\mathrm{ff}}})^{3}
\nonumber\\
    &      
       + f_{3.54544} \, ({a}/{a_{\mathrm{ff}}})^{3.54544}
       + f_{4.54544} \, ({a}/{a_{\mathrm{ff}}})^{4.54544}
 \,,
\label{eq:fit-ansazt-001}
\end{align}
where $f_{0},f_{1},f_{2}$, and $f_{3}$ are independent of $c_{3N}$ while $f_{3.54544}$ and $f_{4.54544}$ depend on $c_{3N}$.

\begin{figure}[!ht]
        \centering
        \begin{minipage}[b]{0.43\textwidth}
                \includegraphics[width=\textwidth]{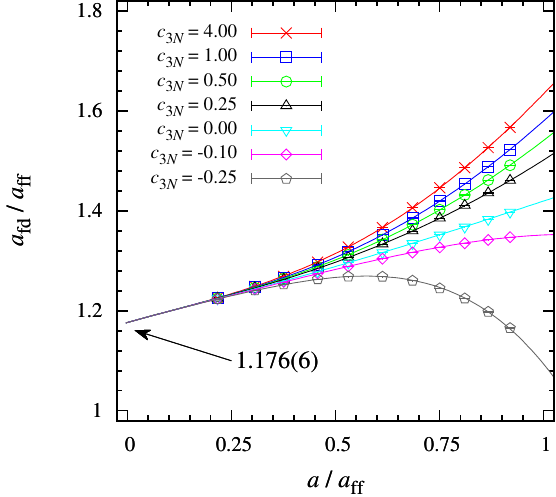}
                \vspace{-0.5cm}
        \end{minipage}
        \hspace{1.0cm}
        \begin{minipage}[b]{0.43\textwidth}
                \includegraphics[width=\textwidth]{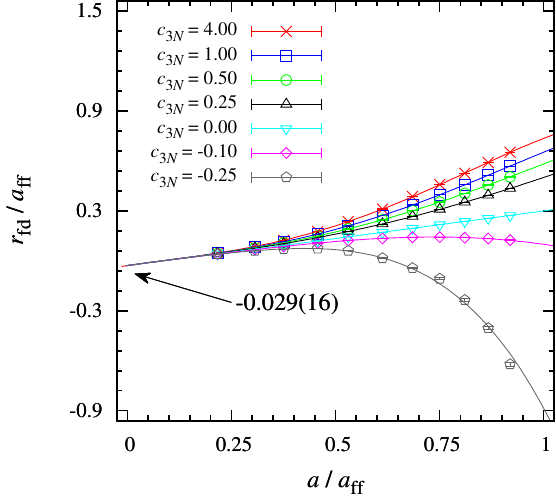}
        \end{minipage}
        \caption{(Left panel) The continuum-limit extrapolation of the fermion-dimer scattering length $a_{\mathrm{fd}}$. (Right panel) The continuum-limit extrapolation of the fermion-dimer effective range $r_{\mathrm{fd}}$. The final results are $a_{\mathrm{fd}}/a_{\mathrm{ff}} = 1.176(6)$ and $r_{\mathrm{fd}}/a_{\mathrm{ff}} = -0.029(16)$.}
        \label{fig:fermiondimer_a0_r0}
\end{figure}
The extrapolation  fits for the scattering length and the effective range are shown in Fig.~\ref{fig:fermiondimer_a0_r0}. The final results for the scattering parameters are $a_{\mathrm{fd}}/a_{\mathrm{ff}}=1.176(6)$ and $r_{\mathrm{fd}}/a_{\mathrm{ff}}=-0.029(16)$, which are in good agreement with semi-analytic continuum calculations $a^{cont.}_{\mathrm{fd}}/a_{\mathrm{ff}} = 1.1791(2)$ and $r^{cont.}_{\mathrm{fd}}/a_{\mathrm{ff}} = -0.0383(3)$~\cite{Bedaque:1997qi,Bedaque:1998mb,Gabbiani:1999yv,Bedaque:2002yg}. The error bars include the uncertainty from the effective range expansion fits and the continuum limit extrapolation. Given the quality of the extrapolation fits in Fig.~\ref{fig:fermiondimer_a0_r0}, we conclude that the three-particle forces make a contribution that is consistent with the conformal scaling prediction of $(a/a_{\rm ff})^{3.54544}$.

\subsection{Dimer-dimer scattering}

We now compute dimer-dimer scattering.   We consider a four-particle system of two spin-up
fermions and two spin-down fermions. Our lattice Hamiltonian
has the form
\begin{equation}
H = H_0 + V_{2N} + V_{3N}+V_{4N},
\end{equation}
where the four-particle interaction is introduced
in Eq.~(\ref{eqn:4NInteraction}). For our analysis we define the parameter $c_{3N,4N}$ and set $C^{(1)}_{3N}=2\,C^{(2)}_{3N}=c_{3N,4N}$  and $C^{(1)}_{4N}=2\,C^{(2)}_{4N}=-3\,c_{3N,4N}$. In the following we quote the value of $c_{3N,4N}$ in
lattice units. 

As in the fermion-dimer calculations, we use the
Lanczos eigenvector method to obtain the finite-volume energies
for the dimer-dimer system for various interaction coefficients $C_{2N}$
and $c_{3N,4N}$ and lattice lengths $L$. From the dimer-dimer energies 
$E^{(L)}_{\mathrm{dd}}$ in the center-of-mass frame, we determine the relative
momentum $p$ using
\begin{align}\label{eq:Edd_L}
E^{(L)}_{\mathrm{dd}} = \frac{p^2}{2\mu^{\ast}_{\mathrm{dd}}}
-2B^{\,}_{\mathrm{d}} - 2\tau_{\mathrm{d}}(\eta) \, \Delta B^{(L)}_{\mathrm{d}}, 
\end{align}
and then use Eq.~(\ref{eqn:Luescher-001}) to extract the dimer-dimer scattering
phase shifts. 

\begin{figure}[!ht]
\includegraphics[width=0.8\textwidth]{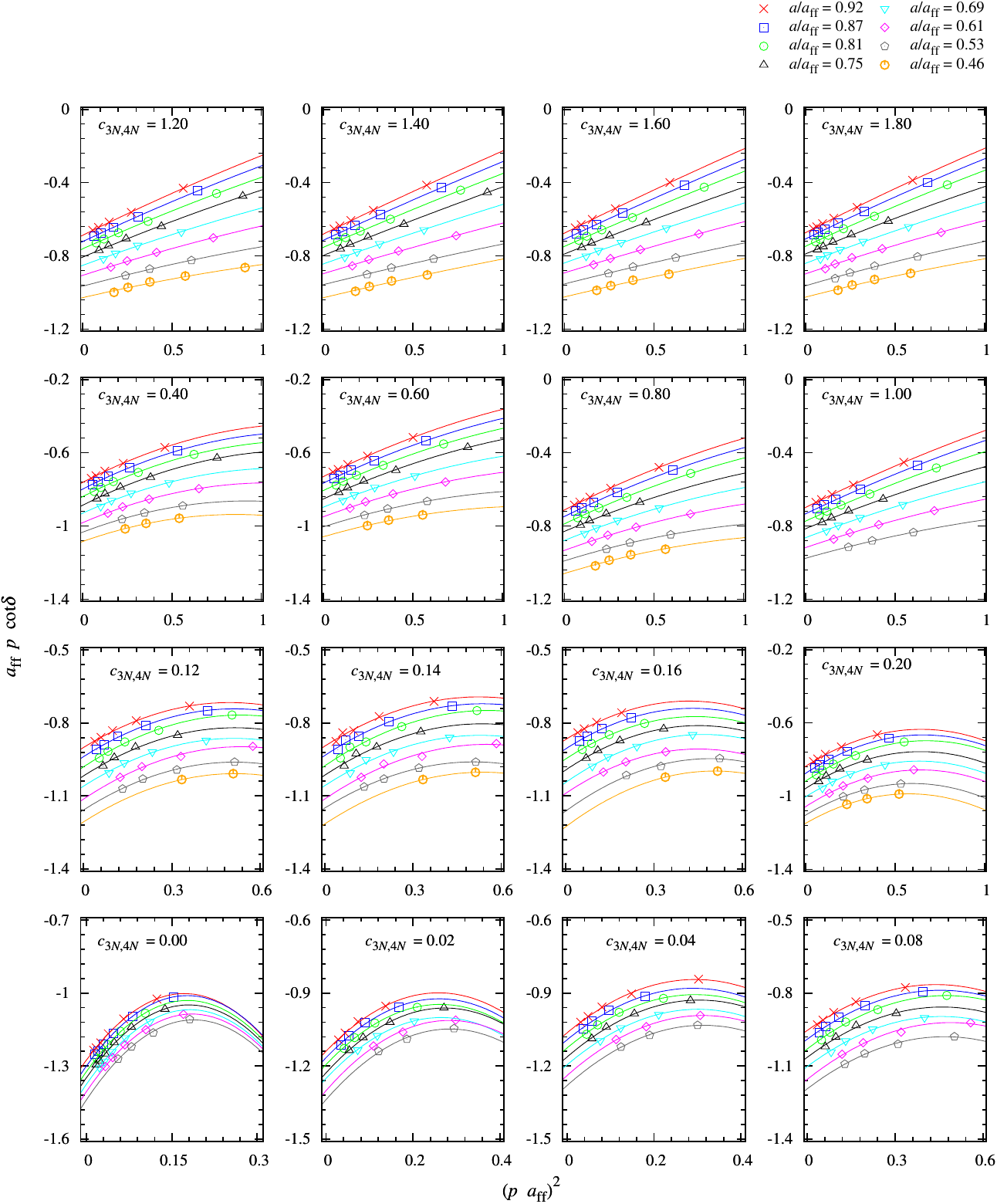}
\caption{The dimer-dimer scattering results for various values of the multi-particle
coupling $c_{3N,4N}$ in lattice units and various ratios of the lattice spacing
$a$ to the fermion-fermion scattering length $a_{\rm ff}$. We plot $a_{\mathrm{ff}}\,p\cot\delta$ versus $(a_{\mathrm{ff}}\,p)^{2}$ in the center-of-mass frame. The points are the lattice data, and the lines are the fits to the effective range expansion.}
\label{fig:dimerdimer_phaseshifts}
\end{figure}

The results for the dimer-dimer phase shifts are shown in Fig.~\ref{fig:dimerdimer_phaseshifts}.
We plot $a_{\mathrm{ff}}\,p\cot\delta$ versus $(a_{\mathrm{ff}}\,p)^{2}$
for various values of the multi-particle
coupling $c_{3N,4N}$ and various ratios of the lattice spacing
$a$ to the fermion-fermion scattering length $a_{\rm ff}$.  The values quoted
for $c_{3N,4N}$ are in lattice units.  In each case we make a fit using the truncated
effective range expansion 
\begin{equation}
a_{\mathrm{ff}}\,p\cot\delta =- \frac{1}{a_{\rm dd}/a_{\rm ff}} + \frac{1}{2}r_{\rm
dd}/a_{\rm ff} \cdot (a_{\rm ff}\,p)^2 + O(p^{4}) ,
\end{equation}
where $a_{\rm dd}$ and $r_{\rm dd}$ are the dimer-dimer scattering length
and  effective range respectively.  We observe in Fig.~\ref{fig:dimerdimer_phaseshifts} that the multi-particle coupling $c_{3N,4N}$  has a stronger impact on the dimer-dimer scattering results than we had seen for $c_{3N}$ in the fermi-dimer scattering results. This is due to finite-volume effects.  It is not possible at present to go to very large volumes in the four-particle system calculations.  The amount of memory required scales as $\ell^9$, where $\ell=L/a$ is the lattice length measured in lattice units. In practice it is difficult to go much beyond $\ell = 12$.
This is in contrast with the three-particle system where the scaling is as
$\ell^6$, and one can reach values of $\ell$ several times larger.

In the absence of the multi-particle
coupling $c_{3N,4N}$, we find that the finite-volume corrections for the four-particle
system are significant.  These findings are consistent with Ref.~\cite{Lee:2006vp},
which discussed a four-particle chain-like excitation wrapping
around the lattice boundaries.
In addition to the leading $(a/a_{\rm ff})^{3.54544}$ dependence on $c_{3N,4N}$, we also have corrections proportional to $(a/a_{\rm ff})^{3.54544}$ times a term proportional to the finite-volume correction of the dimer wave function \cite{Luscher:1985dn},
\begin{equation}
(a/a_{\rm
ff})^{3.54544} \frac{e^{-L/a_{\rm ff}}}{L/a_{\rm ff}}.
\end{equation}
Written in terms of $\ell$, this becomes
\begin{equation}
(a/a_{\rm
ff})^{3.54544} \frac{e^{-\ell a/a_{\rm ff}}}{\ell a/a_{\rm ff}}=(a/a_{\rm
ff})^{2.54544} \frac{e^{-\ell a/a_{\rm ff}}}{\ell},
\end{equation}   
and so for fixed $\ell$ we get  a correction that na{\"i}vely appears to be of a lower order than the expected $(a/a_{\rm
ff})^{3.54544}$ scaling and with a rather complicated dependence on $a/a_{\rm ff}$.  Even though this dependence on $a/a_{\rm
ff}$ is an artificial combination of lattice and finite volume effects, we can still extrapolate the lattice data to obtain the correct continuum limit.

\begin{figure}[!ht]
\centering
\includegraphics[width=0.75\textwidth]{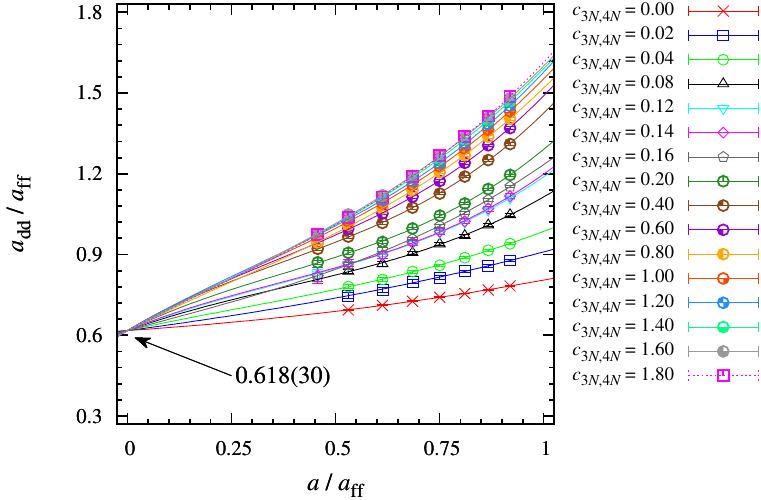}
\caption{The continuum-limit extrapolation of the dimer-dimer
scattering length $a_{\mathrm{dd}}$.  The final result is $a_{\mathrm{dd}}/a_{\mathrm{ff}} = 0.618(30)$.}
\label{fig:dimerdimer_a0}
\end{figure}

In light of the complications from residual finite-volume effects, we use a simpler continuum extrapolation scheme for $a_{\rm dd}$ and $r_{\rm dd}$.  For $a_{\rm dd}$ we use
a simple functional form
\begin{equation}
f({a}/{a_{\mathrm{ff}}}) = f_{0}
       + f_{1} \, ({a}/{a_{\mathrm{ff}}})  
       + f_{2} \, ({a}/{a_{\mathrm{ff}}})^{2}
        + f_{3} \, ({a}/{a_{\mathrm{ff}}})^{3}
\end{equation}
with $f_{0}$ independent of $c_{3N,4N}$, but allowing $f_{1}$, $f_{2}$, and $f_{3}$ to vary with $c_{3N,4N}$. In Fig.~\ref{fig:dimerdimer_a0} we show the continuum-limit extrapolation for the dimer-dimer
scattering length $a_{\mathrm{dd}}$.  The final result we obtain is $a_{\mathrm{dd}}/a_{\mathrm{ff}}
= 0.618(30)$, which is in good agreement with the most accurate determinations in the literature, $a_\mathrm{dd}/a_\mathrm{ff}=0.60\pm0.01$~\cite{Petrov:2004zz,Petrov:2005zz},
$a_\mathrm{dd}/a_\mathrm{ff}=0.605\pm0.005$~\cite{Greene:2009}, and $a_\mathrm{dd}/a_\mathrm{ff}=0.60$~\cite{Aurel:2004}. The error bar includes the uncertainty from the effective range expansion fits and the continuum limit extrapolation.  

As we increase $c_{3N,4N}$, the three-particle interaction becomes more repulsive. This repulsive interaction impedes the formation of the chain-like excitation wrapping around the periodic boundary  which was observed in Ref.~\cite{Lee:2006vp}.  So the finite-volume corrections are smaller for large positive values of $c_{3N,4N}$, and we expect to recover the usual $(a/a_{\rm ff})^{3.54544}$ dependence on $c_{3N,4N}$.  This is consistent with the results in Fig.~\ref{fig:dimerdimer_a0}.  For the largest values of $c_{3N,4N}$, the coefficients of the first and second powers of $a/a_{\rm ff}$ are approximately independent of $c_{3N,4N}$.

\begin{figure}[!ht]
\centering
\includegraphics[width=0.75\textwidth]{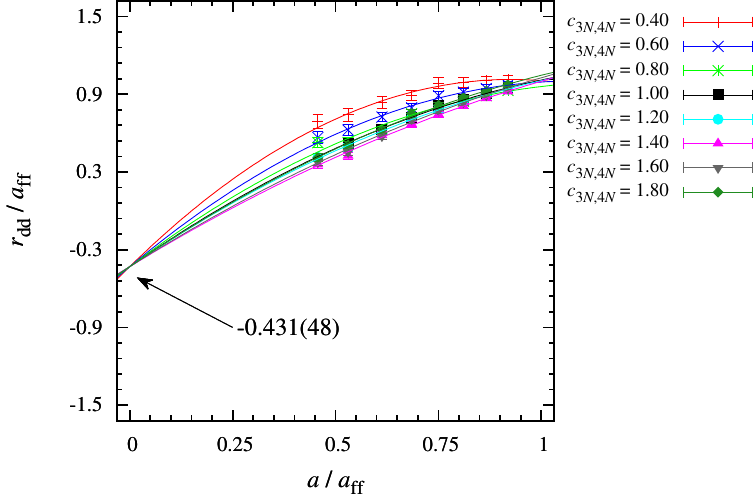}
\caption{The continuum-limit extrapolation of the dimer-dimer
effective range $r_{\mathrm{dd}}$.  The final result is $r_{\mathrm{dd}}/a_{\mathrm{ff}}
= -0.431(48)$.}
\label{fig:dimerdimer_r0}
\end{figure}

For the continuum extrapolation of $r_{\rm dd}$, we use only the data with $c_{3N,4N}
\ge 0.40$, where the three-particle interaction is quite repulsive and finite-volume
effects are small. We use
the functional form\begin{equation}
f({a}/{a_{\mathrm{ff}}}) = f_{0}
       + f_{1} \, ({a}/{a_{\mathrm{ff}}})  
       + f_{2} \, ({a}/{a_{\mathrm{ff}}})^{2}
\end{equation}
with $f_{0}$ independent of $c_{3N,4N}$, but allowing $f_{1}$ and
$f_{2}$ to vary with $c_{3N,4N}$.  In Fig.~\ref{fig:dimerdimer_r0} we show
the continuum-limit extrapolation for the dimer-dimer
effective range $r_{\mathrm{dd}}$.  We note that as $c_{3N,4N}$ increases, the  coefficients
of the first and second
powers of $a/a_{\rm ff}$ are approximately independent of $c_{3N,4N}$. This is consistent with the expected $(a/a_{\rm ff})^{3.54544}$ dependence on $c_{3N,4N}$.  The final result we obtain is $r_{\mathrm{dd}}/a_{\mathrm{ff}} = -0.431(48)$.  The error bar includes the uncertainty from the effective range expansion fits and the
continuum limit extrapolation. This value is different from previous estimates in the literature, $r_\mathrm{dd}/a_\mathrm{ff}
\approx 0.12$ \cite{vonStecher:2007zz} and $r_\mathrm{dd}/a_\mathrm{ff} \sim 2.6$ \cite{Lee:2007ae}.  However in each of the previous estimates, the size of the systematic errors have not been quantified.  In particular the analysis in Ref.~\cite{Lee:2007ae} was plagued by the same large finite-volume effects we have discussed here. In our analysis we used the repulsive three-particle interaction to reduce the size of the finite-volume corrections.


To further analyze the sign of the effective range, we have considered a simple
model of the dimer-dimer system consisting of two fundamental particles.
Due to the Pauli repulsion between identical particles, it is very plausible
that the dimer-dimer interaction has the characteristics of a repulsive Yukawa
interaction at long distances, since this is the functional form of the dimer
wave function~\cite{Petrov:2005zz2}. Therefore, we have considered a dimer-dimer interaction of
the form
\begin{align}
V(r) = V_{G}(r)   \, \theta(R_{g} -r) + V_{S}(r)   \, \theta(r-R_{g})  \,
\theta(R_{y} -r) + V_{Y}(r) \, \theta(r-R_{y}),
\end{align}
where $V_G(r)$ is a Gaussian potential up to radial distance $R_g$, $V_Y (r)$ is a long-range repulsive Yukawa potential starting from radial distance of $R_y$, $V_{S}(r)$ is a cubic spline function, and $\theta$ is a unit step function.  In all cases where the scattering length is positive, we find that the effective range is negative. We find that the repulsive Yukawa potential plays an important role making the effective range negative. These findings support our lattice result of a negative effective range for the dimer-dimer system and is also consistent with the negative value for the fermion-dimer effective range.
 
\section{Summary and Conclusions}
\label{sec_conclusion}

We have used lattice effective field theory to compute the scattering length and effective range of dimer-dimer scattering in the universal limit of large fermion-fermion scattering length.  To benchmark our numerical lattice methods, we first calculated fermion-dimer scattering.  The scattering phase shifts were computed by calculating finite-volume scattering energies and applying L\"uscher's finite-volume method. In our calculations we included a  three-particle interaction in order to generate additional data to be used in the continuum-limit extrapolations.  The dependence on the three-particle interaction coefficient $c_{3N}$ was consistent with the $(a/a_{\rm ff})^{3.54544}$ dependence predicted by conformal scaling in the unitarity limit.  Extrapolating to the continuum limit, we obtained the values $a_{\mathrm{fd}}/a_{\mathrm{ff}}=1.176(6)$ and
$r_{\mathrm{fd}}/a_{\mathrm{ff}}=-0.029(16)$, in excellent agreement with previous calculations of fermion-dimer scattering length and effective
range.  

We then used the same methods to calculate dimer-dimer scattering and extracted the dimer-dimer scattering length $a_\mathrm{dd}$ and effective range $r_\mathrm{dd}$.
In this case we used a multi-particle interaction coefficient $c_{3N,4N}$.  We found that the finite-volume corrections could be reduced by making the three-particle interaction sufficiently repulsive.  Using L\"uscher's finite-volume method, we determined the dimer-dimer scattering phase shifts.  We then extracted the values $a_\mathrm{dd}$ and $r_\mathrm{dd}$ and performed continuum-limit extrapolations.  For the scattering length we obtained $a_{\mathrm{dd}}/a_{\mathrm{ff}} = 0.618(30)$, in good agreement with published results.  For the effective range we found $r_{\mathrm{dd}}/a_{\mathrm{ff}} = -0.431(48)$, which is different from previous estimates. However this new result represents the first calculation with quantified and controlled systematic errors.

Finally, we considered a simple model of the dimer-dimer system as two fundamental particles interacting via a short-range Gaussian interaction and a repulsive Yukawa potential to mimic the Pauli repulsion between identical particles. We found that the effective range is negative for cases where the scattering length is positive.  This may explain why both the dimer-dimer and fermion-dimer effective ranges are negative.

Our results should have immediate applications to the universal physics of shallow dimers.  One particularly useful application is in the determination of the ground-state energy density of a dilute gas of shallow dimers.  The dimers behave as repulsive Bose particles, and the energy density has been determined up to order $\rho\,a_{\rm dd}^{3}$ 
\cite{Lee:1957a,Lee:1957b,Wu:1959,Braaten:1999} as well as the first correction proportional to {\rm $r_{\rm dd}/a_{\rm dd}$ \cite{Braaten:2001a}. This new value for the dimer-dimer effective range suggests that higher-order corrections to the energy density of a dilute gas of shallow dimers could be larger than previously thought.

\begin{acknowledgments}
The authors are grateful for discussions with Yusuke Nishida and acknowledge partial support from the U.S. National Science Foundation  grant No. PHY-1307453,
the DFG (SFB/TR 110, ``Symmetries and the Emergence of Structure in QCD'') 
and the BMBF (contract No.~05P2015 - NUSTAR R\&D).
The work of UGM was also supported in part by The Chinese Academy of Sciences 
(CAS) President's International Fellowship Initiative (PIFI) grant no. 2017VMA0025.
Computing resources were provided by the Higher Performance Computing centers at Mississippi State University, North Carolina State University and RWTH Aachen. 
\end{acknowledgments}


\end{document}